\documentclass[twocolumn]{ceurart}

\usepackage{listings}
\usepackage{algorithm}
\usepackage{algorithmic}

\begin{document}

\copyrightyear{2024}
\copyrightclause{Copyright for this paper by its authors.
  Use permitted under Creative Commons License Attribution 4.0
  International (CC BY 4.0).}

%\conference{Woodstock'22: Symposium on the irreproducible science,
%  June 07--11, 2022, Woodstock, NY}

\title{A Best-of-Both Approach to Improve Match Predictions and Reciprocal Recommendations for Job Search}

\author[1]{Shuhei Goda}[%
%orcid=0009-0005-9656-1420,
email=shu@wantedly.com,
]
\address[1]{Wantedly, Inc., Tokyo, Japan}

\author[1]{Yudai Hayashi}[%
email=yudai@wantedly.com,
]

\author[2]{Yuta Saito}[%
email=ys552@cornell.edu,
]
\address[2]{Cornell University, Ithaca, NY, USA}

\begin{abstract}
Matching users with mutual preferences is a critical aspect of services driven by reciprocal recommendations, such as job search. To produce recommendations in such scenarios, one can predict match probabilities and construct rankings based on these predictions. However, this direct match prediction approach often underperforms due to the extreme sparsity of match labels. Therefore, most existing methods predict preferences separately for each direction (e.g., job seeker to employer and employer to job seeker) and then aggregate the predictions to generate overall matching scores and produce recommendations. However, this typical approach often leads to practical issues, such as biased error propagation between the two models. This paper introduces and demonstrates a novel and practical solution to improve reciprocal recommendations in production by leveraging \textit{pseudo-match scores}. Specifically, our approach generates dense and more directly relevant pseudo-match scores by combining the true match labels, which are accurate but sparse, with relatively inaccurate but dense match predictions. We then train a meta-model to output the final match predictions by minimizing the prediction loss against the pseudo-match scores. Our method can be seen as a \textbf{best-of-both (BoB) approach}, as it combines the high-level ideas of both direct match prediction and the two separate models approach. It also allows for user-specific weights to construct \textit{personalized} pseudo-match scores, achieving even better matching performance through appropriate tuning of the weights. Offline experiments on real-world job search data demonstrate the superior performance of our BoB method, particularly with personalized pseudo-match scores, compared to existing approaches in terms of finding potential matches.
\end{abstract}


\maketitle

\section{Introduction}

Online platforms for job search and dating have increasingly been influencing how people find employment and establish personal connections~\cite{su2022optimizing}. These services rely heavily on the ability to match users effectively via recommendations, as this directly impacts user satisfaction and engagement~\cite{tomita2023fast,tomita2022matching}. To enable users to efficiently find suitable matches, these platforms should recommend user pairs with mutual preferences. However, achieving accurate and efficient matching in these domains presents distinct challenges~\cite{tomita2023fast,kleinerman2018optimally}.

Although standard methods around recommender systems such as matrix factorization~\cite{hu2008collaborative,rendle2012bpr} can be applied using the matches naturally observed on the service as training labels, this approach, which we call the \textit{direct match prediction} approach, often becomes ineffective in real practice, because the match labels are even sparser than typical implicit feedback due to its mutual nature~\cite{Zhao14}. Therefore, existing methods predict preferences separately for each side of the interaction (e.g., job seeker to employer and employer to job seeker) and produce recommendations based on some aggregated scores, e.g., their mere product or harmonic mean~\cite{Pizzato10, Pizzato13, Neve19, Xia15}. While this approach, which we call the \textit{predict-then-aggregate} approach, allows for the utilization of more abundant one-side feedback signals, such as profile views or message exchanges, the aggregation is often done somewhat heuristically. The use of aggregation functions designed without careful consideration of the service can lead to unexpected flaws, such as biased error propagation, resulting in performance lag, as observed in our experiments. 

Particularly, to address the difficult tradeoff between the true match labels (accurate but sparse) and match predictions with some aggregation function (inaccurate but dense), we developed and implemented a novel and simple method based on pseudo-match scores that leverage the strengths of both approaches: the accuracy of true match labels and the dense coverage of match predictions. In particular, our approach generates pseudo-match scores by taking a weighted average of the true match labels and match predictions. These pseudo-match scores then serve as target scores for training a meta-model that learns to directly predict the likelihood of a successful match given a user pair. Importantly, our method also allows for user-specific weights to construct personalized pseudo-match scores, enabling even better matching performance through appropriate tuning of the weights by better capturing the unique characteristics of individual users. Our approach to produce effective reciprocal recommendations though pseudo-match scores have shown to be effective through implementations on production data collected on a real job matching service.

The contributions of the paper can be summarized as follows.
\begin{itemize}
    \item We formulate the problem of reciprocal recommendation from the match prediction perspective.
    \item We develop a novel \textbf{best-of-both} method that leverages \textit{pseudo-match scores} by aggregating true match labels and match predictions, and then learn a meta-model based on these labels to make more accurate match predictions.
    \item We also propose a method for tuning the creation of pseudo-match scores at the user- or segment- level to achieve further improvements.
    \item We demonstrate the effectiveness of the proposed method in production offline experiments, showing that it outperforms typical approaches in terms of matchmaking, particularly for smaller user segments such as relatively inactive users.
\end{itemize}

\section{Related Work}
Reciprocal recommendation systems have garnered significant attention in recent years, particularly in domains where mutual interest between parties is critical, such as online dating, social networking, and job search applications. Unlike traditional recommendation systems, which focus on the preferences of a single party, reciprocal recommendation models account for the preferences of both parties involved, aiming to maximize the likelihood of a successful match. In this section, we review the relevant literature on reciprocal recommendation systems, particularly in the context of job search, and highlight the advancements in this domain.

\subsection{Traditional Recommendation Systems in Job Search}
Traditional recommendation systems in job search applications typically utilize collaborative filtering, content-based filtering, and hybrid approaches to suggest job opportunities to candidates or candidates to recruiters. These systems have been successful in improving job matching efficiency by analyzing user profiles, job descriptions, and historical data to identify potential matches~\citep{guttman1998agent,burke2002hybrid}. Platforms like LinkedIn and Indeed employ recommendation algorithms that leverage user interaction data, such as job clicks, applications, and endorsements, to suggest relevant job postings or candidates~\cite{paparrizos2011machine,malinowski2006matching}. However, these systems often focus primarily on the preferences of one party—typically the job seeker—without sufficiently considering the preferences of the employer, which can result in suboptimal matches~\cite{xia2014predicting}.

\subsection{Reciprocal Recommendation Systems}
Reciprocal recommendation systems extend beyond traditional models by considering the preferences and constraints of both parties involved in a match. These systems have been widely explored in online dating~\citep{pizzato2010recon,xia2015reciprocal}, where mutual interest is crucial for forming a connection. In the job search domain, reciprocal recommendation systems must account for the preferences of both job seekers and employers. For example, a job seeker may prefer a job that aligns with their skills and career aspirations, while an employer seeks candidates with specific qualifications and cultural fit. The challenge lies in balancing these preferences to suggest matches that are mutually satisfactory.

Recent work has expanded on the complexity of modeling these interactions. \citet{yildirim2021bideepfm} and \citet{luo2020rrcn} propose a deep learning-based reciprocal recommendation framework that combines user interaction history with profile matching to enhance prediction accuracy in both job and candidate recommendations. Furthermore, \citet{liu2024linksage} introduced a graph neural network-based approach for reciprocal recommendation in job search applications, where user-job interactions are modeled as a bipartite graph to better capture the underlying relationships between entities.

\subsection{Reciprocal Recommendations in Job Search}
Recent studies have begun exploring the application of reciprocal recommendation systems in job search platforms. \citet{mine2013reciprocal} developed a reciprocal recommendation framework for job matching that incorporates both candidate and employer preferences, using a two-sided matching model to optimize the likelihood of successful employment outcomes. Their work demonstrates the potential of reciprocal systems to improve job or candidate recommendation efficiency by reducing mismatches and increasing the satisfaction of both sides of the users.
Moreover, collaborative filtering techniques tailored to reciprocal settings have been proposed to enhance recommendation accuracy. \citet{zheng2023reciprocal} introduced a collaborative filtering model for job matching that integrates mutual preference information and interaction history between candidates and employers. 

Several recent studies on optimizing recommendations in matching markets have focused on addressing the problem of \textit{congestion} rather than solely on match prediction to further maximize the number of matches. \citet{su2022optimizing} propose a method that optimizes rankings for matching markets by considering that the probability of a job seeker responding to a scout from a company can decrease as the job seeker receives more and more scouts. In the context of online dating, \citet{tomita2022matching} introduce a method to aggregate the preferences of each side based on ideas from matching theory in economics to address the congestion problem. Both works highlight the importance of generating recommendations that effectively leverage match predictions to maximize the number of matches. However, our work focuses more on improving match prediction accuracy, and combining our method with the frameworks of \citet{su2022optimizing} and \citet{tomita2022matching} could be an interesting future direction.

\section{Preliminaries}

\subsection{Problem Formulation}

We first formulate the problem of reciprocal recommendation with a focus on the job search application. Let $c \in \mathcal{C}$ represent a company index and $j \in \mathcal{J}$ represent a job seeker index, where $\mathcal{C}$ and $\mathcal{J}$ are the sets of companies and job seekers available on the platform, respectively.
In the job search problem, there are preferences from both sides: companies have preferences for job seekers, and job seekers have preferences for companies.
We use $p^{c \rightarrow j}$ to denote the preference from company $c$ to job seeker $j$, which can be interpreted as the probability of company $c$ sending an scout to job seeker $j$. Similarly, we define the preference in the other direction, $p^{j \rightarrow c}$, as the probability of job seeker $j$ responding to company $c$.
Using these notations, we can define the probability of observing a successful match between a particular company-job seeker pair $(c, j)$ as $$m(c,j) = p^{c\rightarrow j} \cdot p^{j \rightarrow c}.$$

\begin{figure}
\centering
\includegraphics[width=1.05\linewidth]{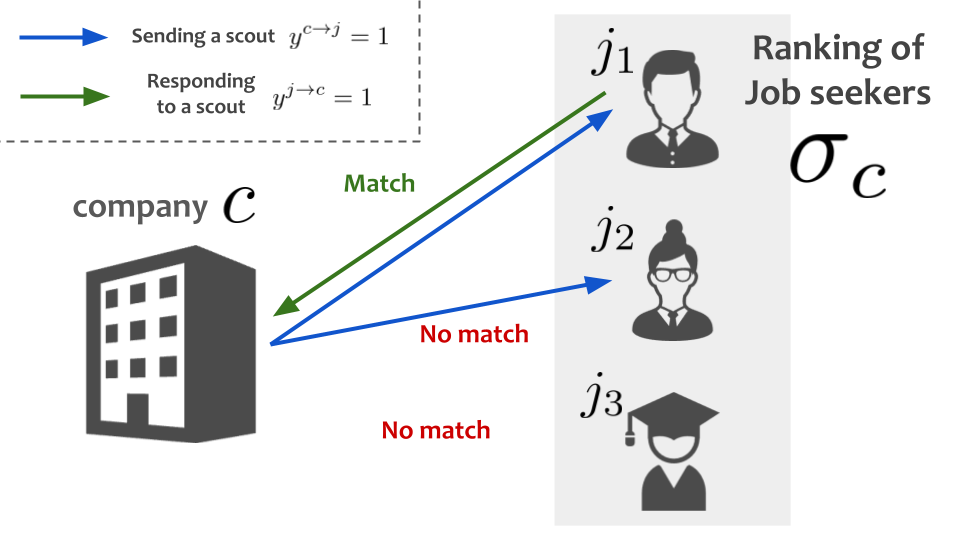}
\caption{The figure illustrates our problem of reciprocal recommendation. The platform generates a ranking of job seekers for companies, and then companies decide whether to send a scout to the job seekers in the ranking. Each job seeker who receives a scout then decides whether to respond. A successful match occurs only when a response from the job seeker is observed.}
\label{fig:problem}
\end{figure}

For the ease of discussion, we focus on the task of producing a recommendation list of job seekers for each company such that the resulting number of matches is maximized, as depicted in Figure~\ref{fig:problem}.\footnote{It is easy to consider the recommendation of the opposite direction, i.e., recommending a list of job postings to job seekers.} 
This can be achieved by presenting the following \textit{optimal} ranking $\sigma_c^*$ to each company $c$:
\begin{align}
    \sigma_c^* := \mathrm{argsort}_{j \in \mathcal{J}} \; m(c,j) \label{eq:optimal}
\end{align}
Note that in our work, we do not address specific problems related to reciprocal recommendations, such as congestion~\cite{su2022optimizing,xia2015reciprocal}, and focus on the match prediction task.

If we can produce the optimal rankings as defined in Eq.~\eqref{eq:optimal} for the companies, we can maximize the number of resulting matches. However, the optimal ranking depends on the true match probabilities, which are not available to us. Therefore, we must first estimate the probabilities $m(c,j)$ for available pairs $(c,j)$ based on training data to feasibly produce rankings of job seekers to present to companies, optimizing the number of matches as much as possible.

\subsection{Typical Approaches}

There exist two high-level approaches to tackle the problem of match predictions, namely \textit{direct match prediction} and \textit{predict-then-aggregate}. The following briefly describes these typical approaches and discusses their limitations.

\subsubsection{Direct Match Prediction (DMP)}
The DMP approach simply predicts the match probabilities based on observed match labels to produce rankings. Suppose we have training data of the following form:
\begin{align}
    \mathcal{D} = \{(c_i, j_i, m_i)\}_{i=1}^n, \label{eq:dmp-data}    
\end{align}
where $i$ is the data index, $n$ is the size of the training dataset, and $m_i \in \{0,1\}$ is a binary match indicator sampled based on the true match probability of the corresponding company-job seeker pair, i.e., $m_i \sim m(c,j)$.

Based on the training dataset in Eq.~\eqref{eq:dmp-data}, DMP learns a prediction model of the match probabilities as
\begin{align}
   \hat{m} = \mathrm{argmin}_{\hat{m}'} \sum_{i=1}^n \ell(\hat{m}' (c_i, j_i),  m_i) ,  \label{eq:dmp}
\end{align}
where $\hat{m}$ is a learned match prediction model, which can be parameterized by any off-the-shelf supervised machine learning methods, such as linear models, matrix factorization, random forests, or neural networks. $\ell(\cdot, \cdot)$ is a loss function such as the cross-entropy loss. Note also that it is straightforward to extend this to cases with company and job seeker features by replacing $c_i$ and $j_i$ in Eq.~\eqref{eq:dmp} with their respective features, $x_{c_i}$ and $x_{j_i}$.

After solving Eq.~\eqref{eq:dmp}, DMP produces the ranking of job seekers for each company as follows:
\begin{align}
    \hat{\sigma}_c^{dmp} := \mathrm{argsort}_{j \in \mathcal{J}} \; \hat{m} (c,j), \label{eq:dmp-ranking}
\end{align}
which aims to maximize the resulting matches based on the prediction model $\hat{m} (c,j)$.

Although this DMP approach is simple and direct in the sense that it directly predicts matches, it often struggles with the sparsity issue when learning a model as in Eq.~\eqref{eq:dmp}. Specifically, in our job search problem, the binary match label $m_i$ is likely to be extremely sparse due to its two-sided nature—actions from both sides must occur for a match to be observed.\footnote{The sparsity of the match label can be rigorously defined as $\sum_i^n m_i / |\mathcal{C}||\mathcal{J}|$.} This extreme sparsity often makes it difficult for the DMP approach to learn an accurate match prediction model, leading to ineffective rankings.

\subsubsection{Predict-then-Aggregate (PtA)}
To address the issue of sparsity, the literature on reciprocal recommendations often employs the PtA approach instead~\cite{tomita2022matching,xia2015reciprocal}. This approach first predicts the preferences from companies to job seekers and from job seekers to companies using two separate models, and then aggregates their predictions to make a final match prediction and ranking.

This approach leverages training data of the following form:
\begin{align}
    \mathcal{D} = \{(c_i, j_i, y_i^{c \rightarrow j}, y_i^{j \rightarrow c})\}_{i=1}^n, \label{eq:pta-data}    
\end{align}
where $y_i^{c \rightarrow j} \in \{0,1\}$ is a binary indicator representing whether an action (e.g., sending a scout) from a company $c$ to a job seeker $j$ is observed, and $y_i^{j \rightarrow c} \in \{0,1\}$ is another binary indicator representing whether an action (e.g., responding to a scout) from a job seeker $j$ to a company $c$ is observed. These indicators are sampled from the corresponding probability distributions, i.e., $ y_i^{c \rightarrow j} \sim p^{c \rightarrow j} $ and $ y_i^{j \rightarrow c} \sim p^{j \rightarrow c} $.

Leveraging the training dataset in Eq.~\eqref{eq:pta-data}, PtA obtains two separate prediction models to predict the actions of each side as follows.
\begin{align}
   &\hat{p}^{c \rightarrow j} = \mathrm{argmin}_{\hat{p}'} \sum_{i=1}^n \ell(\hat{p}' (c_i, j_i),  y_i^{c \rightarrow j}) \notag ,\\
   &\hat{p}^{j \rightarrow c} = \mathrm{argmin}_{\hat{p}'} \sum_{i=1}^n \ell(\hat{p}' (c_i, j_i),  y_i^{j \rightarrow c}) ,
   \label{eq:pta}
\end{align}
where $\hat{p}^{c \rightarrow j}$ and $\hat{p}^{j \rightarrow c}$ are learned prediction models for each label, which can be parameterized by any off-the-shelf supervised machine learning methods, as in the DMP approach.

After solving the two prediction tasks in Eq.~\eqref{eq:pta}, PtA produces the ranking of job seekers for each company as
\begin{align}
    \hat{\sigma}_c^{pta} := \mathrm{argsort}_{j \in \mathcal{J}} \; M (\hat{p}^{c \rightarrow j}, \hat{p}^{j \rightarrow c}) , \label{eq:pta-ranking}
\end{align}
which aims to maximize the resulting matches based on the prediction models $\hat{p}^{c \rightarrow j}$ and $\hat{p}^{j \rightarrow c}$. $M(\cdot,\cdot)$ is often referred to as an aggregation function, whose role is to combine the two separate predictions. The simplest choice is the product (i.e., $M(x,y) = x \cdot y$), but other functions, such as the harmonic mean, are sometimes used as well.

Although this PtA approach addresses the sparsity issue of DMP by dividing the match prediction problem into two separate prediction tasks with relatively dense binary labels, it often struggles with a significant challenge due to the substantial difference between the probabilities $p^{c \rightarrow j}$ and $p^{j \rightarrow c}$. Frequently, the interaction that occurs first (in our case, $p^{c \rightarrow j}$) is denser than the subsequent interaction (in our case, $p^{j \rightarrow c}$). As a result, the first model is likely to exhibit much greater variation in its predictions, and errors from this model can significantly impact the final ranking performance. This limitation of the PtA approach, due to the use of two independent models, leaves much room for improvement in match-making effectiveness, as we will demonstrate in the following sections.

\section{The \textit{Best-of-Both} (BoB) Approach}
The core of our method to improve reciprocal recommendations generates pseudo-match scores by appropriately combining the following two types of feedback.

\begin{itemize}
    \item \textbf{True match labels}, which indicate whether a successful match has been observed between the two sides of the users. They are highly accurate as they represent real match outcomes, and are used by the DMP approach. However, they are extremely sparse because it needs positive interactions from both sides to occur. \vspace{2mm}
    \item \textbf{Match predictions}, which are generated based on existing recommendation methods based on the PtA approach. These predictions estimate the likelihood of a positive action (such as sending a scout and responding to it) from one side to the other and will be aggregated. While being less accurate than the true match labels, they provide denser feedback.
\end{itemize}

Match predictions are typically used to make reciprocal recommendations (i.e., the PtA approach) because true match labels are often too sparse to learn a model by itself.
However, it is true that these two sources of information retain completely the opposite properties regarding their density and accuracy, so it would be possible to construct a better approach by effectively combining them. 
We simply achieve this by generating pseudo-match scores, which are defined by the weighted average of the true match labels and match predictions. 
Specifically, the pseudo-match score for a pair $(c, j)$ is calculated as follows:
\begin{equation}
    s_{pseudo}(c, j;\alpha) = \alpha \cdot \underbrace{m}_{\textit{true match label}} 
    + (1 - \alpha) \cdot \underbrace{\left(\hat{p}^{c \rightarrow j} \cdot \hat{p}^{j \rightarrow c} \right)}_{\textit{match prediction}},  \label{eq:pseudo-match scores}
\end{equation}
where $m$ represents the true match label (1 for a successful match, 0 otherwise), $\hat{p}^{c \rightarrow j}$ and $\hat{p}^{j \rightarrow c} \in [0, 1]$ represent the predictions given by two separate models trained as in Eq.~\eqref{eq:pta}, and $\alpha \in [0, 1]$ is a parameter that controls the relative importance of the true match labels versus the match predictions. The parameter $\alpha$ provides flexibility in balancing the contributions of the true match labels and match predictions. A higher value of $\alpha$ gives more weight to the true match labels, while a lower value of $\alpha$ places more emphasis on the match predictions. The optimal value of $\alpha$ may vary depending on the characteristics of the dataset and, therefore, needs to be tuned based on cross-validation.

While the approach described above uses a global $\alpha$ parameter for all users, we can further improve the method by introducing \textit{personalized} weights. Specifically, for each pair $(c,j)$, we define a personalized weight $\alpha_{c,j}$ to calculate the personalized pseudo-match score as
\begin{equation}
    s_{pseudo}(c, j;\alpha_{c,j}) = \alpha_{c,j} \cdot m(c, j) + (1 - \alpha_{c,j}) \cdot \hat{p}^{c \rightarrow j} \cdot \hat{p}^{j \rightarrow c},  \label{eq:personalized-pseudo-match scores}
\end{equation}
where $\alpha_{c,j}$ is the weight uniquely defined for the pair $(c,j)$. The personalized weights allow the model to adapt to different user types, such as users with varying levels of activity. In practice, it is also possible to perform the personalization of the weights at the user segment level rather than the individual user level and assign an $\alpha$ value specific to each segment such as new and old users.

After defining the pseudo-match scores, we use them as a prediction target for training a meta-model to directly predict the probability of a potential match for $(c, j)$. This can be done as
\begin{equation} 
   \hat{f} = \mathrm{argmin}_{f'} \sum_{(c, j)} \ell(f'(c, j), s_{pseudo}(c, j;\alpha)) ,  \label{eq:meta}
\end{equation}
where $f$ is the meta-model. Once we have trained the meta-model, we make recommendations of job seekers ($j$) to the company side ($c$), as 
\begin{align}
    \sigma_c^{bob} = {\operatorname{argsort}}_j\, f (c, j)  \label{eq:proposed}
\end{align}
where $\sigma_c$ is the ranking of job seekers induced by $f$, and it will be presented to company $c$.
Algorithm~\ref{algo:bob} provides the pseudo-code of our proposed BoB approach.

There are several advantages to our BoB method over the typical approaches described in the previous section.
First, it does not rely solely on the true match labels ($m$) and is therefore more robust to the sparsity issue compared to the DMP approach.
Second, it also does not depend exclusively on the two separate prediction models ($\hat{p}^{c \rightarrow j}, \hat{p}^{j \rightarrow c}$), and by incorporating the true match labels, it avoids the issues associated with the PtA approach.
Third, by personalizing the weight $\alpha$ as in Eq.~\eqref{eq:personalized-pseudo-match scores}, our method can flexibly prioritize either the DMP or PtA approaches depending on their effectiveness for each individual user.
The following section empirically demonstrates these advantages of our method on a real-world dataset.

\begin{algorithm}[t]
\caption{The BoB Approach (our proposed method)}
\label{algo:bob}
\begin{algorithmic}[1]
\REQUIRE training data $\mathcal{D}$; weights; $\alpha_{c,j}$
\ENSURE a learned meta model $\hat{f}$
\STATE Perform the PtA approach as in Eq.~\eqref{eq:pta} and obtain two separate models $\hat{p}^{c \rightarrow j}$ and $\hat{p}^{j \rightarrow c}$
\STATE Calculate the pseudo-match scores by using the true match labels, match predictions, and given weights $\alpha_{c,j}$, as in Eq.~\eqref{eq:personalized-pseudo-match scores}
\STATE Train a meta-model to predict the pseudo-match scores as in Eq.~\eqref{eq:meta}
\end{algorithmic}
\end{algorithm}

\section{Experiment on Production Job Seekers Recommendation Data}

To evaluate the effectiveness of our proposed BoB method, we conduct offline experiments on a production dataset collected from a real-world job search platform. Specifically, we aim to answer the following research questions through these experiments: \vspace{2mm}

\textbf{RQ1.} Does using both true match labels and match predictions improve matching performance compared to baseline methods? \vspace{2mm}

\textbf{RQ2.} Do optimal $\alpha$ values vary across different user segments? \vspace{2mm}

\textbf{RQ3.} Does personalization of $\alpha$ values lead to better matching performance compared to non-personalized approaches? \vspace{2mm}

\subsection{Dataset Description}

The real-world production data used for our experiments were collected from the job search platform \textit{Wantedly Visit}\footnote{\url{https://www.wantedly.com}}. Wantedly Visit is a service in Japan that connects job seekers with companies and includes a scouting feature where companies can send scouts to job seekers to arrange casual interviews. In this experiment, a match ($m(c, j) = 1$) is defined as a company sending a scout to a job seeker and the job seeker responding positively. Non-matches ($m(c, j) = 0$) occur when this mutual interaction does not take place.

The dataset covers the period from November 2023 to February 2024. The features used for model training and inference are derived from two main sources: profile information and action logs. The profile information for job seekers includes their educational background, job category, work history, self-introduction, future aspirations, and skills. For companies, it includes company information and job posting details. Additionally, we utilize action logs from both job seekers and company recruiters on the platform.

\subsection{Compared Methods}

To rigorously define the baseline methods, we use the following notations.

\begin{itemize}
\item {$\hat{p}^{c \rightarrow j}$}: prediction of the preference from company $c$ to job seeker $j$.
\item {$\hat{p}^{j \rightarrow c}$}: prediction of the preference from job seeker $j$ to company $c$.
\item {$\hat{m}(c, j) = M(\hat{p}^{c \rightarrow j}(c, j), \hat{p}^{j \rightarrow c}(c, j))$}: prediction of the match between $c$ and $j$ based on an aggregation function $M$.
\end{itemize}

We compare our proposed method with the following baseline approaches, all of which can be expressed with corresponding aggregation function $M(\hat{p}^{c \rightarrow j}, \hat{p}^{j \rightarrow c})$:

\begin{itemize}
    \item \textbf{Scout-Only}: $M(\hat{p}^{c \rightarrow j}, \hat{p}^{j \rightarrow c}) = \hat{p}^{c \rightarrow j}$.
    \item \textbf{Reply-Only}: $M(\hat{p}^{c \rightarrow j}, \hat{p}^{j \rightarrow c}) = \hat{p}^{j \rightarrow c}$.
    \item \textbf{Multiplication}: $M(\hat{p}^{c \rightarrow j}, \hat{p}^{j \rightarrow c}) = \hat{p}^{c \rightarrow j} \cdot \hat{p}^{j \rightarrow c}$.
    \item \textbf{Harmonic Mean}: $M(\hat{p}^{c \rightarrow j}, \hat{p}^{j \rightarrow c}) = \frac{2 \hat{p}^{c \rightarrow j} \cdot \hat{p}^{j \rightarrow c}}{\hat{p}^{c \rightarrow j} + \hat{p}^{j \rightarrow c}}$.
\end{itemize}

\subsection{Detailed Experimental Design}

To implement our method, we adopt the Gradient Boosting Decision Tree~\cite{Jerome01} as the meta-model $f$ and train it using Eq.~\eqref{eq:meta}. We investigate the effectiveness of five different global $\alpha$ values: 0.0, 0.25, 0.5, 0.75, and 1.0.
To further assess the effectiveness of our BoB method with personalized $\alpha$ values, we define three segments (High, Middle, Low) based on the activity level of the companies. Then, for each segment, we explore four different $\alpha$ values: 0.0, 0.25, 0.5, and 0.75. Note that we exclude 1.0 from the personalized exploration, as our experience indicates that it consistently leads to performance degradation.
We evaluate and compare the NDCG@10 of the methods calculated with the true match labels in the test dataset derived from real recommendation logs collected on Wantedly Visit. Note that we use a time-based split to create the training and test datasets and employ 5-fold cross-validation to tune hyperparameters. It is also important to note that our offline evaluation pipeline has been validated for consistency with the online experiment results on the platform.

\subsection{Experimental Results}

\subsubsection{RQ1. Does using both true match labels and match predictions improve matching performance compared to baseline methods?}

Table \ref{tab:freq} presents the experimental results. We first observe that Scout-Only and Reply-Only are highly ineffective, underscoring the importance of considering preferences from both sides to identify potential matches on our platform. The table also shows that our method, with a global $\alpha=0.25$, achieves an NDCG@10 score of 0.1021, which is higher than the best baseline (Harmonic Mean) at 0.0979. This indicates that appropriately combining true match labels with match predictions to generate pseudo-match scores leads to improved recommendations by facilitating more matches. We also observe that the effectiveness of the proposed method varies with different $\alpha$ values, highlighting the importance of careful tuning based on validation.

\begin{table}
\caption{Comparison of the effectiveness in reciprocal recommendations between our proposed BoB method and baselines.}
\label{tab:freq}
\scalebox{1.}{
\begin{tabular}{ccl}
\toprule
Methods & NDCG@10 (on test data)\\
\midrule\midrule
Scout-Only & 0.0592 \\
Reply-Only & 0.0886 \\
Multiplication & 0.0969 \\
Harmonic Mean & 0.0979 \\ \midrule
BoB (w/ Global $\alpha=0.00$) & 0.1017 \\
\textbf{BoB (w/ Global $\alpha=0.25$)} & \textbf{0.1021} \\
BoB (w/ Global $\alpha=0.50$) & 0.0926 \\
BoB (w/ Global $\alpha=0.75$) & 0.0944 \\
BoB (w/ Global $\alpha=1.00$) & 0.0932 \\ \midrule
\textbf{BoB (w/ Personalized $\alpha$)} & \textbf{0.1050} \\
\bottomrule
\end{tabular}}
\end{table}

\begin{figure*}
\centering
\includegraphics[width=\textwidth]{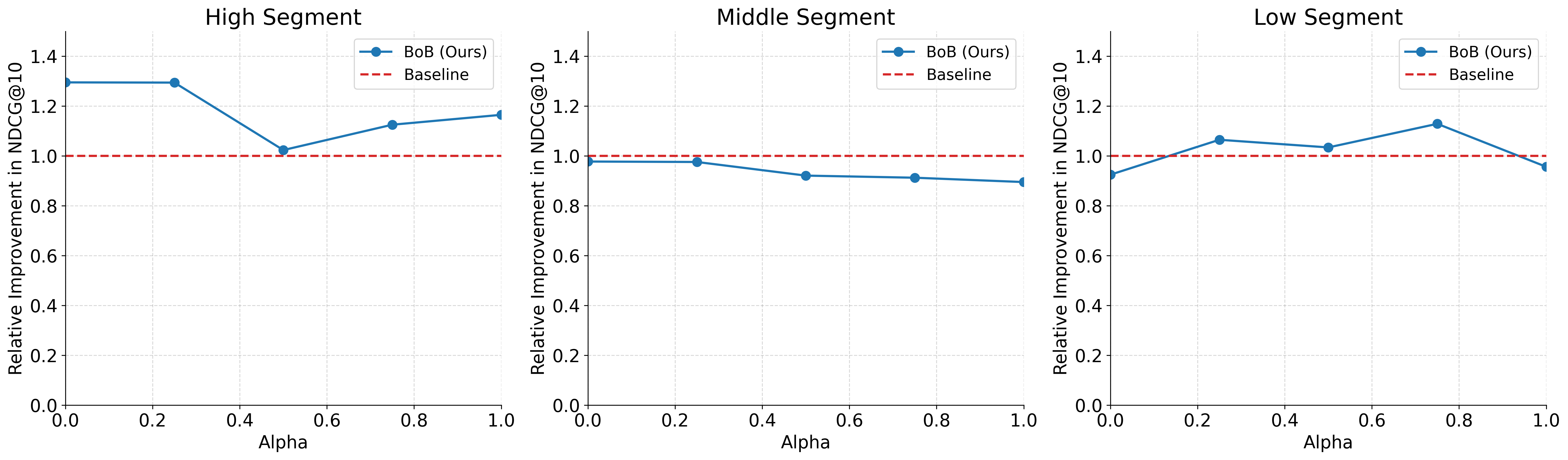}
\caption{Performance comparison of the BoB method versus the best baseline (Harmonic Mean) for varying $\alpha$ values for each segment. The graphs illustrate the relative improvement in NDCG@10 for the High, Middle, and Low activity segments. The x-axis represents $\alpha$ values ranging from 0 to 1, while the y-axis shows the relative performance compared to the best baseline (represented by the horizontal dashed line at 1.0).}
\label{fig:segment_performance}
\end{figure*}

\subsubsection{RQ2. Do optimal $\alpha$ values vary across different user segments?}

To investigate how the optimal $\alpha$ values vary across different user segments, we analyzed the NDCG@10 scores for each company segment at different $\alpha$ values, as shown in Figure \ref{fig:segment_performance}.

The results reveal significant variations in the optimal $\alpha$ values across different segments:

\begin{itemize}
    \item \textbf{High Activity Segment}: This segment shows substantial improvement over the baseline, with lower $\alpha$ values (0.0 and 0.25) yielding the best performance. This suggests that for highly active companies, relying more on match predictions is beneficial. \vspace{1mm}
    \item \textbf{Middle Activity Segment}: This segment shows a slight preference for lower $\alpha$ values (0.0 and 0.25), similar to the High Activity Segment. However, even at these optimal $\alpha$ values, performance is marginally worse than the baseline. \vspace{1mm}
    \item \textbf{Low Activity Segment}: Interestingly, this segment performs best with a high $\alpha$ value (0.75), indicating that true match labels are more valuable for companies with low activity levels.
\end{itemize}

These findings highlight the importance of segment-specific $\alpha$ tuning. The stark contrast in optimal $\alpha$ values between the High Activity and Low Activity segments suggests that the effectiveness of match predictions versus true labels varies significantly across user segments, further justifying the need for a personalized strategy to define $\alpha$.

\subsubsection{RQ3. Does personalization of $\alpha$ values lead to better matching performance compared to non-personalized approaches?}

To evaluate the effectiveness of personalization, we compared the performance of our proposed method with personalized $\alpha$ values against the non-personalized approach. As shown in Table \ref{tab:freq}, the personalized approach achieved an NDCG@10 score of 0.1050, outperforming both the best baseline (Harmonic Mean at 0.0979) and the best non-personalized approach (Global $\alpha$=0.25 at 0.1021).
The optimal $\alpha$ values for the (High, Middle, Low) activity segments were 0.0, 0.75, and 0.75, respectively. These results align closely with our observations from RQ2.
By tailoring the balance between true match labels and match predictions to each segment's characteristics, we can achieve superior matching performance compared to a one-size-fits-all approach.

\subsection{Discussion}

The High Activity segment showed substantial improvement over the baseline, particularly with lower $\alpha$ values (0.0 and 0.25). This suggests that for highly active companies, relying more on match predictions is beneficial. Several factors could explain this phenomenon. First, the use of a meta-model may reduce the impact of error propagation that typically occurs during the aggregation phase in traditional PtA methods. Additionally, the abundance of true labels for this segment might create an ensemble-like effect when combined with the various labels used in pseudo-label creation. Most importantly, the large volume of data for this segment likely results in high-quality match predictions. These labels are both dense and accurate, making them highly valuable for training. The combination of these factors—error mitigation, ensemble effects, and high-quality match predictions—may work synergistically to produce the observed performance improvements in the High Activity segment.

Conversely, the Low Activity segment performed best with a high $\alpha$ value (0.75), indicating that true match labels are more valuable for companies with low activity levels. This could be attributed to several factors. Primarily, the baseline predictions (match predictions) for this segment may be of very low quality due to sparse data. Not only are match labels sparse for this segment, but apply and reply labels are likely scarce as well, leading to poor prediction accuracy for individual preferences. In such situations, the traditional approach of predicting each side of the interaction separately and then aggregating may introduce significant errors. Consequently, directly optimizing for matches using true labels, despite their sparsity, appears to be more effective than relying on potentially inaccurate match predictions. This suggests that when data is limited and predictions are less reliable, leveraging the high accuracy of true match labels, even if they are few in number, yields better results than attempting to use lower-quality, but more abundant, predicted labels.

Interestingly, the Middle Activity segment showed no improvement over the baseline and, in fact, experienced slight performance degradation. This could be because the model predictions for this segment, while denser than true match labels, are not as accurate. At the same time, the true match labels, while more accurate, are not as abundant as in the High Activity segment. As a result, the model struggles to effectively leverage the strengths of both types of feedback. The match predictions may not be of sufficient quality to provide additional valuable information beyond what the true labels offer, yet the true labels may not be numerous enough to compensate for the inaccuracies in the match predictions.

\section{Conclusion and Future Work}

This paper introduces a novel and practical approach to improving reciprocal recommendations for person-job matching using pseudo-match scores. Our BoB approach effectively combines the strengths of true match labels and match predictions, leading to enhanced matching performance. The personalized version of our method further improves performance by adapting to the characteristics of different users or user groups.

While our method has shown promising results, there are several avenues for future exploration. First, it would be interesting to investigate the application of our BoB approach with personalized weights to other domains where reciprocal recommendations are crucial, such as online dating or mentor-mentee matching platforms. Additionally, we aim to explore more sophisticated personalization techniques, including dynamic adjustment of $\alpha$ values based on real-time user behavior and feedback.

\bibliography{workshop}

\end{document}